\documentclass[aps,prl,showpacs]{revtex4}
\usepackage{graphicx}
\usepackage{bm}

\newcommand{\lapr}{La$_{0.625-y}$Pr$_{y}$Ca$_{0.375}$MnO$_{3}$ }

\begin{document}
\setlength{\baselineskip}{2.5\baselineskip}
\title{Scaled frequency-dependent transport in the mesoscopically
    phase-separated colossal magnetoresistive manganite
    La$_{0.625-y}$Pr$_y$Ca$_{0.375}$MnO$_3$}
\author{S. Chaudhuri}
\author{R. C. Budhani}\email{rcb@iitk.ac.in}
 \affiliation{Condensed Matter - Low Dimensional Systems Laboratory\\
Department of Physics, Indian Institute of Technology Kanpur, Kanpur
- 208016, India }
\author{Jiaqing He}
\author{Yimei Zhu}
\affiliation  {Department of Nanoscience, Brookhaven National
Laboratory, Upton, NY-11973}
\date{\today}

\begin{abstract}
We address the issue of massive phase separation (PS) in manganite
family of doped Mott insulators through ac conductivity measurements
on La$_{0.625-y}$Pr$_{y}$Ca$_{0.375}$MnO$_{3}$ (0.375 $\leq$ y
$\leq$ 0.275), and establish applicability of the scaling theory of
percolation in the critical regime of the PS. Measurements of dc
resistivity, magnetization (M(T)) and electron diffraction show
incomplete growth of a ferromagnetic (FM) metallic component on
cooling the high temperature charge ordered (CO) phase well below
Curie temperature. The impedance $\mid$Z(T,f)$\mid$ measured over a
frequency (f) range of 10 Hz to 10 MHz in the critical regime
follows a universal scaling of the form $\approx$
R(T,0)g(f$\xi^{2+\theta}$ ) with $\theta$ $\approx$ 0.86 and the
normalized correlation length varying from 1 to 4, suggesting
anomalous diffusion of holes in percolating FM clusters.
\end{abstract}


\maketitle

The importance of phase separation (PS) in understanding  the
physics of manganite is vital \cite{moreo,mathur}. Systems with
strong electron correlations, like manganites, tend to be
intrinsically phase segregated into regions with different electron
density. This coexistence of energetically near degenerate phases
differing markedly in their properties makes manganites very
susceptible to external perturbations like electric, magnetic and
photon fields and pressure, thereby giving rise to the observed
colossal responses \cite{salamon,tokura,coey,rao}. One interesting
doped manganite in which the phase separation has been imaged with a
variety of techniques is \lapr (LPCMO). The end members of this
pentanary compound, La$_{1-x}$Ca$_{x}$MnO$_{3}$ (LCMO) and
Pr$_{1-x}$Ca$_{x}$MnO$_{3}$ (PCMO), are a ferromagnetic metal (FM)
and a charge ordered (CO) insulator  respectively for the same
doping level of 0.3 $\leq$ x $\leq$ 0.5 \cite{salamon,tokura}. Thus,
in LPCMO even for a fixed doping (x) a transition from the FM to CO
phase can be achieved by tuning the La/Pr ratio. X-ray diffraction
\cite{kiryukhin} and optical studies \cite{lee} on LPCMO single
crystals indicate that above the metal-insulator (MI) transition
temperature T$_{MI}$, there also exists a charge disordered (CD)
insulating phase apart from the CO phase. Cheong and coworkers
\cite{kiryukhin} further proposed that the percolation like MI
transition seen in LPCMO is via the growth of the FM domains at the
expense of the CD phase below the charge ordering temperature
(T$_{CO}$). Magneto-optical (MO) images of crystals of LPCMO (y =
0.3), a related compound, revealed inhomogeneous magnetization and
current distribution over a length scale of several micrometers due
to phase separation \cite{tokunaga}. Submicrometric phase separation
in La$_{0.33}$Pr$_{0.34}$Ca$_{0.33}$MnO$_{3}$  thin films has also
been seen with scanning probe microscopy (SPM) techniques
\cite{zhang}. Using electron diffraction and dark field imaging,
Uehara \textit{et al.} \cite{uehara} obtained phase separation of CO
and FM domains over a scale of $\approx$ 0.5 $\mu$m at low
temperatures. Thus, the PS as observed by most of these real space
imaging techniques is over a micrometer length scale  \cite{loudon}.
At such a length the FM and CO clusters are large enough to treat
electron transport through the system classically. Here we apply the
scaling theory of percolation for the first time to ac conductivity
in a  manganite which shows phase separation below a critical
temperature T$_{MI}$. While this problem has been addressed
theoretically by Mayr \textit{et. al.}  \cite{mayr} in the framework
of a random resistor model, scaling analysis of the  frequency
dependent conductivity which allows determination of the percolation
correlation length $\xi$ has been lacking although a scaling between
dc conductivity and magnetization has been shown for the system
Pr$_{0.63}$Ca$_{0.37}$MnO$_{3}$  \cite{hardy}. In a classical
metal-dielectric percolating system the correlation length $\xi$
diverges as $\xi$ $\sim$ (\textit{p} - \textit{p}$_{c}$)$^\nu$ as
the metal concentration (p) approaches the percolation threshold
(\textit{p}$_{c}$) . For \textit{p} $>$\textit{p}$_{c}$, the dc
conductivity $\sigma$$_{dc}$ scales as (\textit{p} -
\textit{p}$_{c}$)$^\mu$ $\sim$ $\xi^{\mu/\nu}$ and the probability
of belonging to the infinite conducting cluster is P$_{\infty}$
$\sim$ (\textit{p} - \textit{p}$_{c}$)$^\beta$, where $\mu$, $\nu$
and $\beta$ are the critical exponents \cite{zallen}. It has been
pointed out that on a length scale (L) \textit{b} $\ll$ L $\ll$
$\xi$ ( where \textit{b} is the microscopic lattice distance ) the
geometry of percolating clusters is self-similar and on such a scale
all physical measurements are expected to reflect self-similarity
and therefore the bulk properties should be independent of $\xi$
\cite{gefen}. One can then estimate the relative phase fraction (PF)
of the coexisting phases from the value of $\xi$. Although the PF in
certain manganites has been inferred from the SPM measurements, the
surface sensitive nature of these techniques may not allow correct
determination of the PF. In this letter we first establish phase
separation in epitaxial
 films of \lapr indirectly through quantitative measurement of
 magnetization and then present direct imaging of the FM and CO
 domains with Lorentz and dark-field electron microscopy respectively. The ground
 state of this system is a metal with nearly equal FM and CO phase
 fractions. The FM component (\textit{p}) diminishes  with the gain of CO phase on approaching T$_{MI}$. The frequency dependent impedance scales as the
 function g(f$\xi^{2+\theta}$) with $\theta$ $\approx$ 0.86 suggesting anomalous diffusion in
 percolating clusters.


Thin films of thickness $\approx$ 200 nm of \lapr \ with y = 0.275,
0.3, 0.325, 0.35, 0.375 were grown  on (110) oriented single
crystals of NdGaO$_{3}$ (orthorhombic with d$_{110}$ = 3.8595 {\AA})
using pulsed laser deposition (PLD) technique. A KrF excimer laser
operated at 10 Hz with an areal energy density of 2 J/cm$^{2}$/pulse
on the surface of a stoichiometric sintered target of LPCMO was used
for ablation. The oxygen pressure and the substrate temperature
during growth were held at $\approx$ 300 mTorr and 750 $^{O}$C
respectively. Measurements of dc and ac  resistivity were carried
out on 1 mm wide strips of the film in a four probe configuration
using current source, voltmeter and HP-4192A impedance analyzer,
while magnetization was measured with a SQUID magnetometer. For
measurement of DC resistivity, the test current was 10$\mu$A. With
the sample resistance of 2 K$\Omega$ at $\sim$ 140 K, the Joule
heating is $\sim$ 0.2 $\mu$W. Similarly, for AC measurements, the
power dissipation in the samples remains below 100 $\mu$W. The I-V
curves, measured at several temperatures upto a peak current of
$\sim$ 5 mA, remained linear and nonhysteretic indicating the
absence of  any heating effects. For magnetic imaging and
diffraction, we used the JEM2100F-M, a field emission 200 kV
electron microscope with a special custom made field-free objective
lens (residual field $<$ 4Oe). In-situ cooling and heating
experiments were conducted using a Gatan cooling stage filled with
liquid helium and/or nitrogen.  TEM samples were prepared via
standard mechanical thinning and polishing techniques to remove the
substrate.  The final thinning of the films was done using a
Fischione ion miller with a single low voltage argon source at
glancing angles and at liquid nitrogen temperature.


The temperature dependence of the resistivity $\rho(T)$, normalized
to its value   at 280 K   of the sample with  y = 0.275, 0.3, 0.325,
0.35, 0.375 is shown in Fig. 1.{\label{rt}} The $\rho(T)$ is marked
by a semiconductor-like behavior on cooling below room temperature
followed by an insulator-metal [M-I] transition at T$_{MI}$ on
further cooling, which reveals  superheating and supercooling
effects indicating a first-order transition in the system. The
critical temperature is also a monotonically decreasing function of
y as shown in Fig. 1(inset a), in agreement with the reported
behavior of \lapr bulk samples \cite{uehara}. However, unlike some
other CO manganites where a distinct step in the $\rho$ vs T plot at
T $>$ T$_{MI}$ marks  the onset of charge ordering (T$_{CO}$), we
only see a change in the slope of $\rho(T)$ curve as illustrated in
the inset b of Fig. 1. The T$_{CO}$ defined as the onset of a sharp
change in the slope of the $\frac{{d\rho }}{{dT}}$ vs T curve is
$\approx$ 210 K, which is in agreement with our TEM results
discussed in the later part of this paper.

The magnetization (M(T,H)) signal of these films is overwhelmed by
the strong paramagnetic contribution of the Nd$^{3+}$ ions of  the
substrate (NGO) which makes a quantitative measurement of
magnetization erroneous by as much as 50$\%$.
 Nevertheless, the field dependence of magnetization at various temperatures was carefully
extracted  for the y = 0.275 sample and is displayed in Fig. 2. The
onset temperature of spontaneous magnetization(T$_{Curie}$) of this
sample is $\approx$ 190 K, which matches with the insulator-to-metal
transition seen in $\rho(T)$ on cooling the sample [Fig. 1].  The
temperature dependence of saturation moment plotted in the inset of
Fig. 2, shows a much steeper rise and a wrong slope at lower
temperatures, which is not expected from the spin wave renormalized
mean-field dependence of M(T). This observation suggests a
percolative growth of the ferromagnetic phase fraction at T $<$
T$_{MI}$. It needs to be pointed out here that inspite of a large
error in the measurement of the absolute moment, its value per Mn
ion is well below the moment that would result if all Mn$^{3+}$ and
Mn$^{4+}$ site spins were aligned ferromagnetically. This
observation, which indicates incomplete conversion of the CO phase
into the FM phase is supported unequivocally by  our electron
microscopy results. The Lorentz TEM images of the sample taken at 90
K and shown in Fig. 3(a,b) reveal ferromagnetic domains when viewed
along the [010] direction in the P$_{nma}$ setting. The horizontal
lines in the figure are the twin boundaries associated with the
(101) twinning originating from cubic-to-orthorhombic transition at
high temperature. In Lorentz-Fresnel mode, black and white contrast
of divergent and convergent magnetic domain-walls, respectively, can
be seen under underfocus and overfocus imaging conditions
\cite{lloyd}. There are two sets, primary and secondary, of magnetic
domains in the sample. The A and B in Fig. 3 mark a pair of the
180$^{0}$ primary domain walls, with magnetization anti-parallel
across the walls. The secondary domains are those with the 90$^{0}$
domain walls that coincide with the twin boundaries which are likely
to arrest domain motion and result in the large coercive field
($\approx$ 500 Oe) seen in Fig. 2. Fig. 3(c) is a schematic, showing
the local magnetization. It is important to point out that such FM
domains are
 seen sparsely in the sample indicating that only a small
  fraction of the sample volume is ordered ferromagnetically.

To understand the CO and FM phase separation and phase evolution,
the samples were imaged at several temperatures between 15 K and 300
K in dark-field using the ($\frac{1}{2}$ 0 2) or (2 0 $\frac{1}{2}$)
 superlattice reflection associated with the charge
ordering. Fig. 4 shows clearly the evolution and growth of CO
domains at various temperatures  (a) 73 K, (b) 100 K, (c) 115 K, (d)
135 K, (e) 146 K, and (f) 160 K. Since in this temperature regime,
twins exist, they are superimposed with CO.  The increasing volume
fraction of the CO domains is also reflected by the brightness of
the superlattice reflections in the (010)$^{\ast}$ zone diffraction
patterns, as shown in the insets of Fig. 4(a and f). While the
limited area of view in dark-field imaging makes it difficult to
measure the exact volume fraction of the CO and FM domains as a
function of temperature, our estimate yields $\approx$ 50 $\%$ of
each phase at $\approx$ 70 K which remains almost constant below 70
K. The sample appears fully charge ordered between 170 K and 210 K,
and above 210 K it becomes paramagnetic as indicated by the
disappearance of the satellite reflection.

 Having established unequivocally a massive PS in the system, we now come to the key issue of ac-transport.  In Fig. 5(a) we show the
frequency (\textit{f}\hspace {0.5mm}) dependence of impedance of the
y = 0.275 sample at various temperatures. For T = 139 K, which is
below T$_{MI}$ (T$_{MI}$ $\approx$ 175 K for this sample), the FM
fraction is large and \textit{p} $\gg$ \textit{p$_{c}$}. The
isothermal impedance corresponding to  139 K first remains constant
till 10$^{5}$ Hz and then falls off above it as $ \approx$
$\textit{f}\hspace {1mm}^{-x}$ (x $\approx$ 1.8). Some of these
features of ac conductivity have been seen earlier in related
systems \cite{pandey,mercone}. As the T$_{MI}$ is approached, the
critical frequency ($\textit{f}\hspace {0.5mm}\hspace
{0.5mm}^{\ast}$) of deviation from a constant value decreases with
temperature for T $<$ T$_{MI}$ and then follows a reverse trend at T
$>$ T$_{MI}$. This anomalous behavior of the ac-response in the
vicinity of T$_{MI}$ is a characteristic feature of percolating
systems \cite{gefen}, which has  been explained either by using the
ideas of intercluster polarization \cite{imry} in which more of the
dangling ends of metallic clusters contribute to conduction at
higher frequency, or due to delay time effects on holes whose
diffusion length L$_{\textit{f}\hspace {1mm}}$ over the fractal
network of the PS medium scales as L$_{\textit{f}\hspace {1mm}}$
$\sim$$\textit{f}\hspace {1mm}^{-1/(\theta+2)}$, where $\theta =
(\mu-\beta)/\nu$ = 0.8 in 2-dimensions \cite{gefen}. Based on the
approach of anomalous diffusion, the frequency dependence of the
sample impedance has been modeled as \cite{laibowitz},
\begin{displaymath}
{|Z (T,\textit{f}\hspace {1mm})|  =  R (T,0)g(f\xi^{2+\theta})}
\end{displaymath}
The different impedance isotherms at T$_{MI}$ $<$ T $<$ T$_{MI}$
with frequency shown in Fig. 5(a) can be fitted to an universal
graph  shown in Fig. 5(b). This fit was achieved by scaling the
frequency as {f$\xi^{2+\theta}$} for different temperatures. At each
temperature the value of $\xi$ was chosen to give the best fit to
the curve. The correlation length $\xi$ represents the mean spanning
length of the ferromagnetic clusters in the critical regime. It is
expressed in the units of a microscopic length scale `b', which
could be the size ($\approx$ 12 $\AA$) of magnetic polarons in these
systems as seen in small angle neutron scattering experiments
\cite{teresa}. As shown in the inset, the variation of $\xi$ and
resistivity with temperature has the same nature. Similar results
were obtained for y =  0.325 and 0.375 samples.

In summary we have measured the frequency dependence of impedance of
a doped manganite at 3/8 filling (x = 3/8 ) in the critical regime
of metal-insulator phase transition from CO to FM state. High
resolution electron microscopy combined with Lorentz  imaging, which
establish mesoscale phase separation in the material  suggest a
percolative character of this M-I transition. The scaling behaviour
of $\mid$Z(T,$\textit{f }$)$\mid$ in the critical regime reinforces
this conclusion and allows estimation of the mean spanning length of
ferromagnetic clusters in the vicinity of T$_{MI}$.

This work has been supported by internal funding of IIT-Kanpur and
the office of Basic Energy Sciences, U. S. Department of Energy.

\newpage

\begin{figure}
\includegraphics [width=14cm]{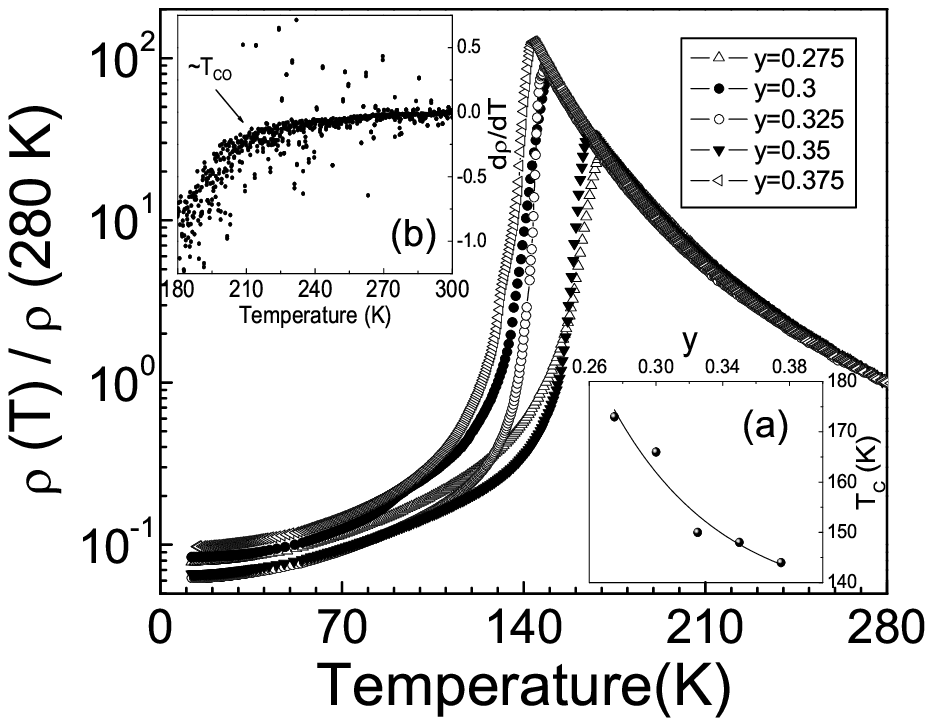}
\caption{ Temperature dependence of resistivity normalized to its
value at 280 K for samples of different y all measured during
heating cycle. Inset `b' shows the variation of $\frac{{d\rho
}}{{dT}}$ with T. The compositional dependence of T$_{c}$ is shown
in inset `a'. }
\end{figure}

\begin{figure}
\includegraphics [width=13cm]{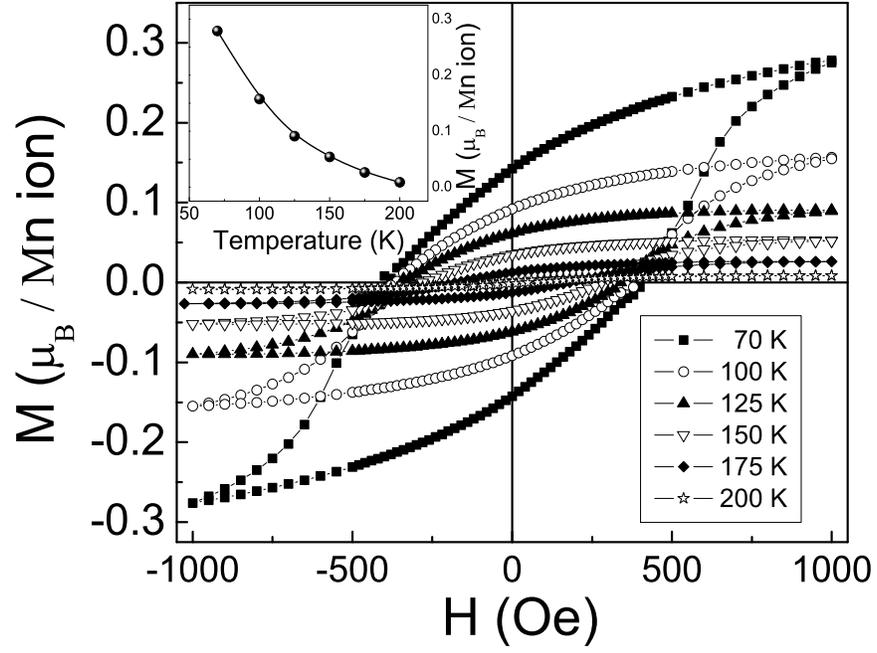}
\caption{ Magnetization vs Field measurements for the y = 0.275
sample at various temperatures. Inset shows the saturation
magnetization as a function of temperature. The absolute value of
the magnetic moment is uncertain by $\approx$ 50\%. }
\end{figure}

\begin{figure}
\includegraphics [width=13cm]{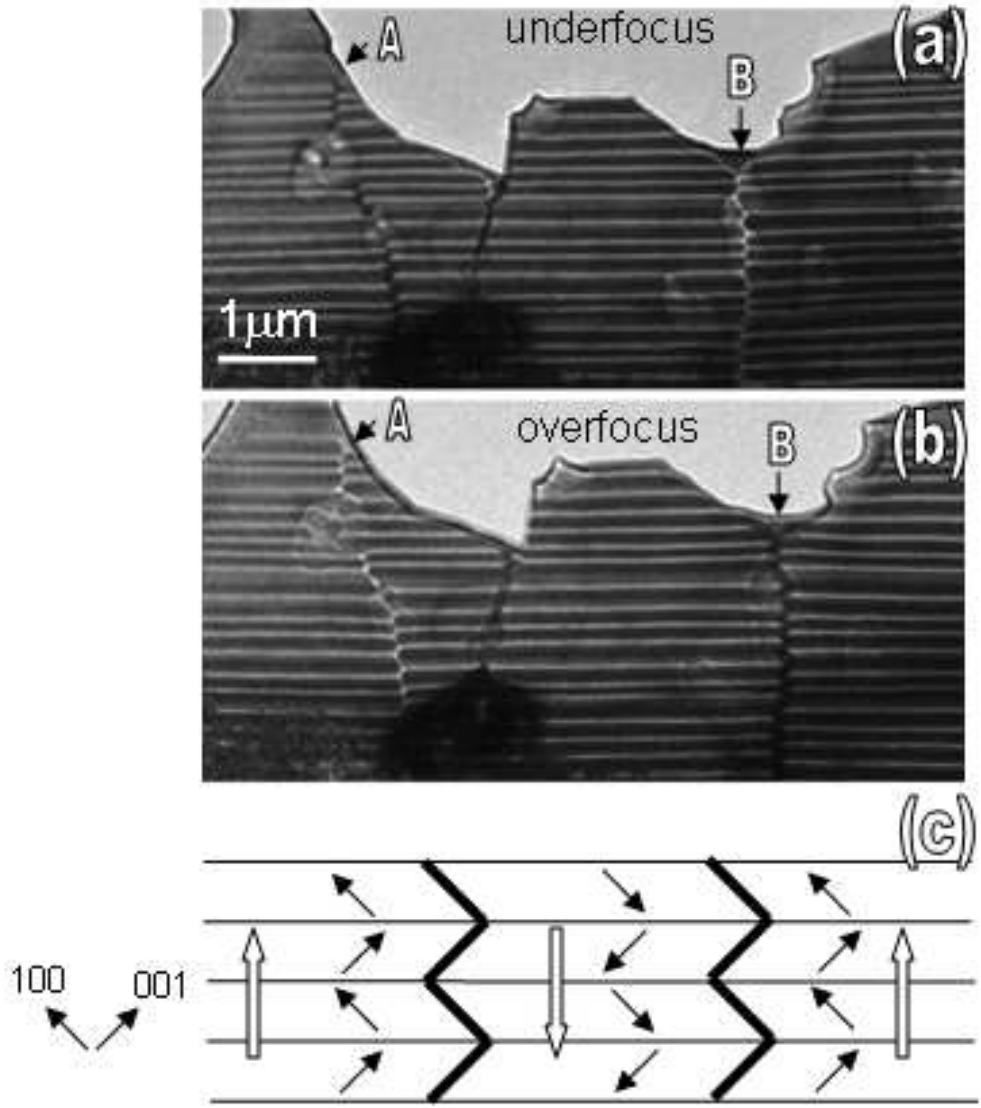}
\caption{Lorentz TEM images of \lapr (y = 0.275) at 90 K.   (a)
underfocus and (b) overfocus image. (c) Schematic drawing of the
local magnetic structure.  The large and small arrows represent the
magnetization of the  primary (180$^{0}$) and secondary (90$^{0}$)
magnetic domains, respectively.  Note, the secondary ferromagnetic
domains coincide with the twinning structure of the sample.  }
\end{figure}

\begin{figure}
\includegraphics [width=16cm]{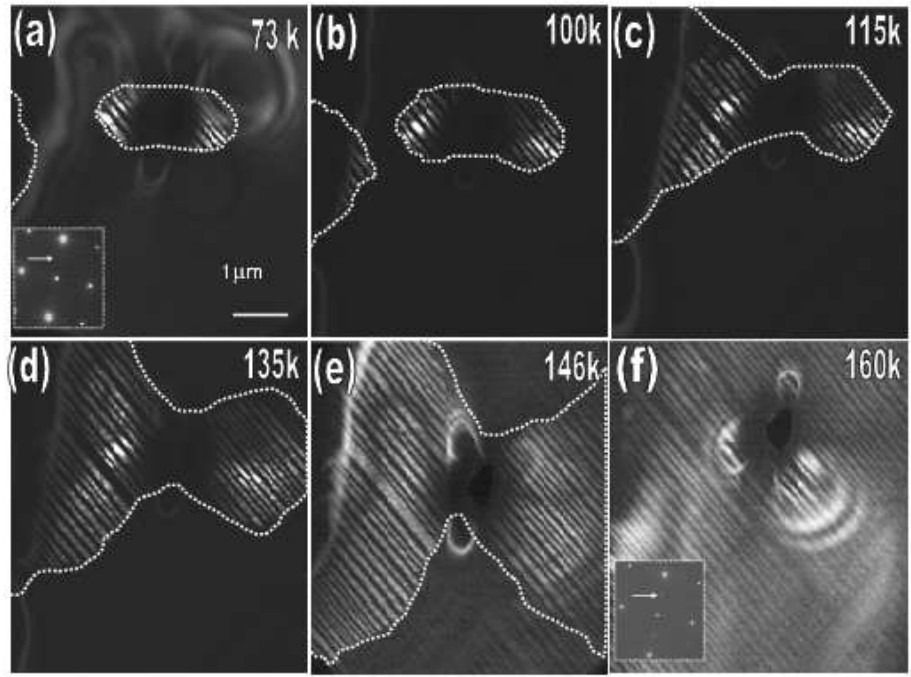}
\caption{Dark-field images of \lapr  (y = 0.275) recorded using the
($\frac{1}{2}$ 0 2) or (2 0 $\frac{1}{2}$) superlattice reflection
showing the evolution and growth of charge ordered domains at
various temperatures. (a) 73 K, (b) 100 K, (c) 115 K, (d) 135 K, (e)
146 K, and (f) 160 K.  For clarity, the walls of the charge-ordered
domains are outlined.  The (010)$^{\ast}$ zone diffraction patterns
at 73K and 160K are also included (see insets). Note, the
significant difference in intensity of the superlattice reflection
at the two temperatures. }
\end{figure}

\newpage

\begin{figure}
\includegraphics [width=9cm]{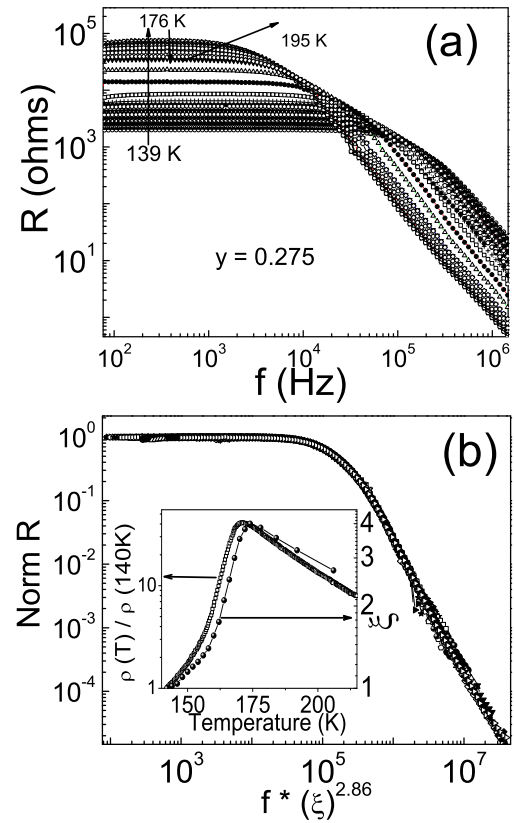}
\caption{(a) Frequency dependence of impedance of y = 0.275 sample at
different temperatures. (b) The universal curve generated form the
isotherm impedance shown in `a' after scaling. The inset of (b) shows
the value of $\xi$ corresponding to different temperatures (right
scale). The  normalized resistivity is also shown (left scale).}
\end{figure}

 \end{document}